\documentclass[
    ,final            
  ]
  {aipproc}

\layoutstyle{6x9}

\usepackage{epsfig}
\def\lsim{\raise0.3ex\hbox{$<$\kern-0.75em\raise-1.1ex\hbox{$\sim$}}}
\def\gsim{\raise0.3ex\hbox{$>$\kern-0.75em\raise-1.1ex\hbox{$\sim$}}}
\begin{document}
\title{Lattice QCD at High Temperature and the QGP}


\author{Frithjof Karsch}{
address={Physics Department, Brookhaven National Laboratory, Upton, NY 11973, 
USA}
}

\begin{abstract}
We review recent progress in studies of bulk thermodynamics of strongly 
interacting
matter, present results on the QCD equation of state and discuss the
status of studies of the phase diagram at non-vanishing quark chemical 
potential.
\end{abstract}

\maketitle

\vspace{-6.7cm}
\hfill {BNL-NT-06/2}
\vspace{5.2cm}

\section{Introduction}

During recent years our knowledge of the thermodynamics of strongly
interacting elementary particles greatly advanced. Lattice
calculations now allow to study also the thermodynamics at non-zero
quark chemical potential ($\mu_q$). The different approaches developed
for this purpose  \cite{Fodor1,us1,Gavai1,Lombardo1,Philipsen1}
are still limited to the regime of high temperature
and small values of the chemical potential, $T\; \gsim\; 0.9T_c,~\mu_q/T\; 
\lsim\; 1$. 
They, however, allow already to analyze the density dependence of the 
equation of state
in a regime relevant for a wide range of energies explored by heavy ion 
experiments and may even be sufficient to establish or rule out the 
existence of a second order phase transition point in the QCD phase diagram.
The existence of such a critical point as endpoint of a line 
of first order phase transition that separates at low temperature 
the low and high density regions in the QCD phase diagram,
is suggested by many phenomenological models.
For small values of $\mu_q/T$ lattice calculations suggest that the 
transition from low to high temperature is not a phase transition; 
the transition during which bulk thermodynamic quantities, e.g. the energy
density, change rapidly and the chiral condensate drops suddenly, is  
a continuous, rapid crossover. It thus 
has been speculated \cite{Stephanov}
that a $2^{nd}$ order phase transition point
exists somewhere in the interior of the QCD 
phase diagram.

\begin{figure}
  \includegraphics[height=.25\textheight]{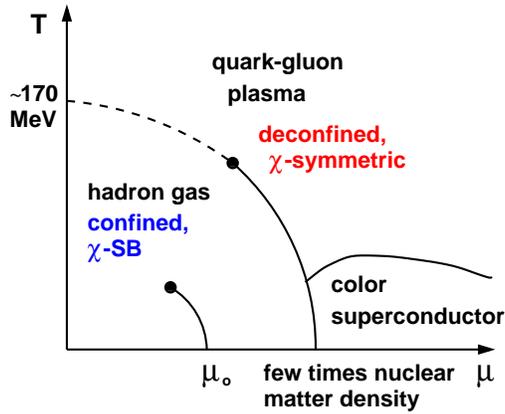}
  \caption{Sketch of the QCD phase diagram.}
\label{fig:phasediagram}
\end{figure}

The generic form of the QCD phase diagram is shown in 
Fig.~\ref{fig:phasediagram}. Although this
phase diagram is well motivated by model calculations little
is known quantitatively about it from lattice calculations. 
All calculations performed so far for QCD with two light
quarks with or without the inclusion of a heavier strange quark
find a smooth but rapid crossover from the low to high temperature regime.
However, these calculations also did not provide evidence for the expected
universal critical behavior in the light quark mass limit, which for 2-flavor
QCD should be that of a $3$-dimensional $O(4)$ or $O(2)$ spin model. 
It generally is expected that calculations closer to the continuum limit are 
needed to unreveal these universal features of the QCD transition. 
In view of this missing evidence for $O(N)$ scaling, it has been argued 
recently that the transition in 2-flavor QCD could also be
a weak first order transition \cite{DiGiacomo}, in which case the entire
transition line in the $\mu$-$T$ phase diagram could be a line of first
order transitions.
 
We will start our survey of lattice calculations on QCD thermodynamics
in the next Section by discussing recent studies of the QCD 
equation of state  at vanishing and non-vanishing chemical potential. 
Section 3 is devoted to a  
discussion of the status of studies of the $2^{nd}$ order phase transition point
({\it chiral critical point}) in the QCD phase diagram.

\section{The QCD equation of state}

\subsection{Vanishing baryon number density}

Most information on the structure of the high temperature phase of QCD and 
the nature of the transition itself has been obtained through lattice 
calculations performed in the limit of vanishing
baryon number density or vanishing quark chemical potential ($\mu_q=0$). 
This limit is most relevant for
our understanding of the evolution of the early universe and  
also is the regime which can be studied experimentally in heavy ion 
collisions at RHIC (BNL) and soon also at the LHC (CERN). The 
experimental accessibility of this regime of dense  
matter also drives the desire to go beyond a qualitative analysis of the 
QCD phase transition and to aim at a numerically accurate determination of 
basic parameters that characterize the thermodynamics of dense 
matter at high temperature.

\begin{figure}
\epsfig{file=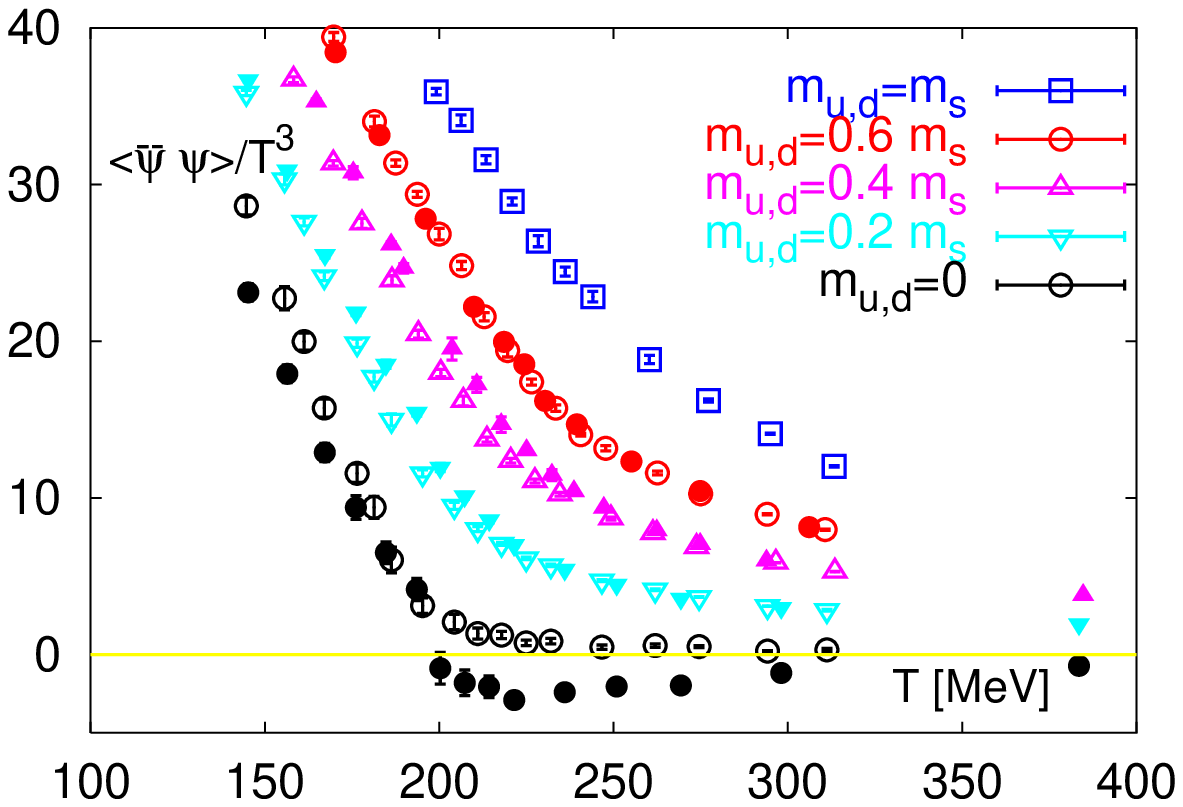,width=64mm}
\epsfig{file=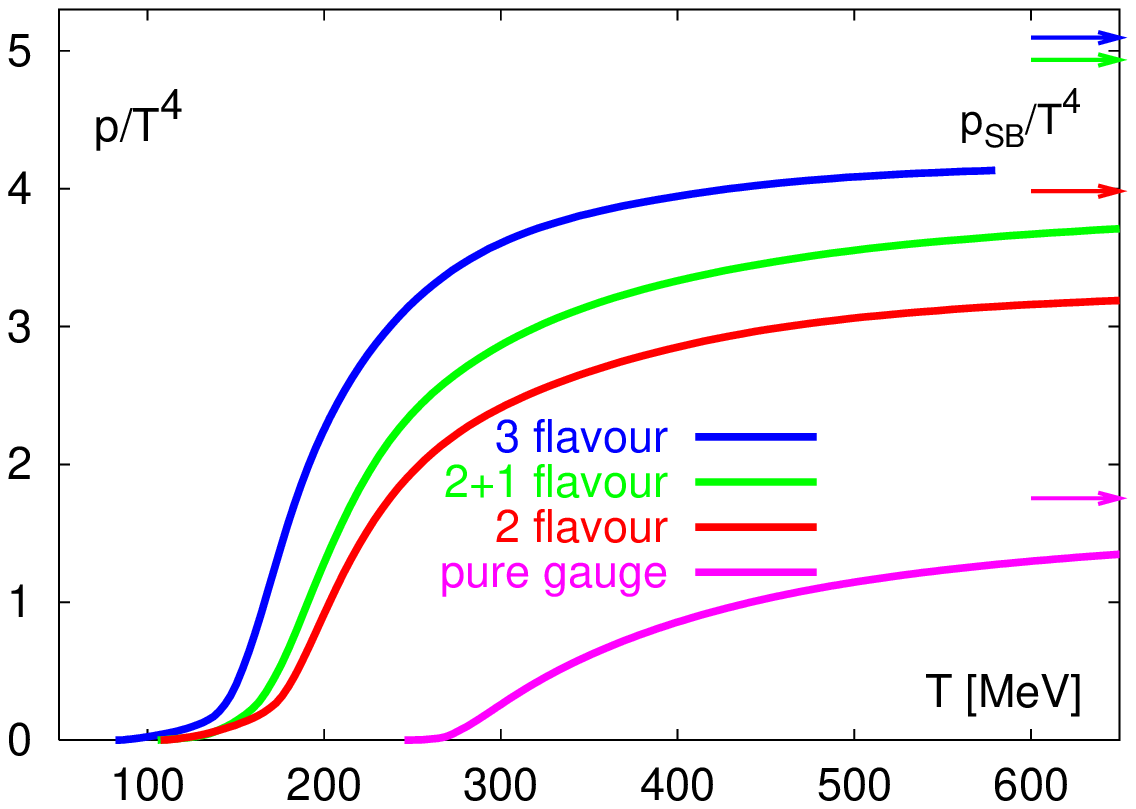,width=64mm}
\caption{\label{fig:chiral} The light quark chiral condensate in QCD
with 2 light up, down and a heavier strange quark mass (open symbols)
and in 3-flavor QCD with degenerate quark masses
(full symbols) \cite{Bernard04}. The right hand part of the figure shows the
pressure calculated in QCD with different number of flavors as well as
in a pure gauge theory \cite{peikert}. Note that (2+1)-flavor QCD here
refers to QCD with two light quarks and a heavier (strange) quark with a mass
proportional to the temperature, $m_s\sim T$.
}
\end{figure}

The transition to the high temperature phase of QCD is related 
to the restoration of chiral symmetry as well as deconfinement; 
the vanishing of the chiral condensate at the transition temperature $T_0$
and the sudden liberation of quark and gluon degrees of freedom is
clearly visible in Fig.~\ref{fig:chiral}.
As can be seen the temperature and quark mass dependence
of the light quark chiral condensate in QCD with two light and a heavier
strange quark is similar to that of QCD with 3 degenerate (light) quarks.
In the high temperature phase, on the other hand, bulk thermodynamic
observables, e.g. the pressure shown in Fig.~\ref{fig:chiral}(left), 
clearly reflect the number of light degrees of freedom and also
are sensitive to the heavier strange quark mass\footnote{In the calculation 
of the 
pressure of (2+1)-flavor QCD shown in Fig.~\ref{fig:chiral}(right) the strange 
quark mass has actually been taken to be proportional to the temperature, 
{\it i.e.} $m_s/T=$const. If one keeps instead $m_s$ fixed to its physical 
value the strange quark effectively becomes light in units of the temperature
and the pressure will gradually approach the high temperature limit of 
3-flavor QCD.}. 

Recent studies of the equation of state concentrated on 
calculations with an almost realistic quark mass spectrum
performed along lines of constant 
physics, {\it i.e.} with light and strange quark masses fixed  
in units of hadron masses rather than the temperature \cite{aoki,Bernard}. 
To reduce cut-off effects induced by the finite lattice spacing 
larger temporal lattices ($N_\tau=6$) and so-called fat links have been used.
At present these calculations still
have been performed with rather small spatial volumes, $V^{1/3}T\simeq 2$.
Close to $T_0$, where the correlation length becomes 
large, a more detailed analysis of the approach to the thermodynamic limit
thus still has to be performed in the future. 

\begin{figure}
  \includegraphics[height=.24\textheight]{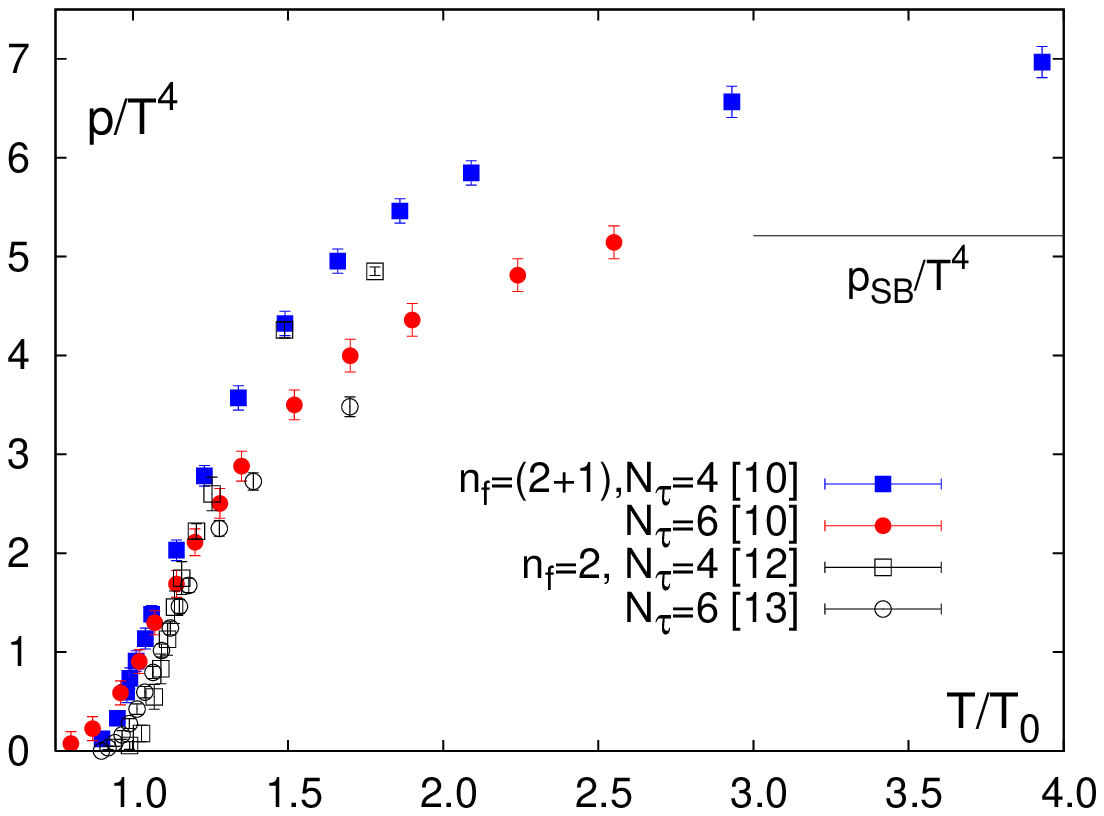}
  \includegraphics[height=.23\textheight]{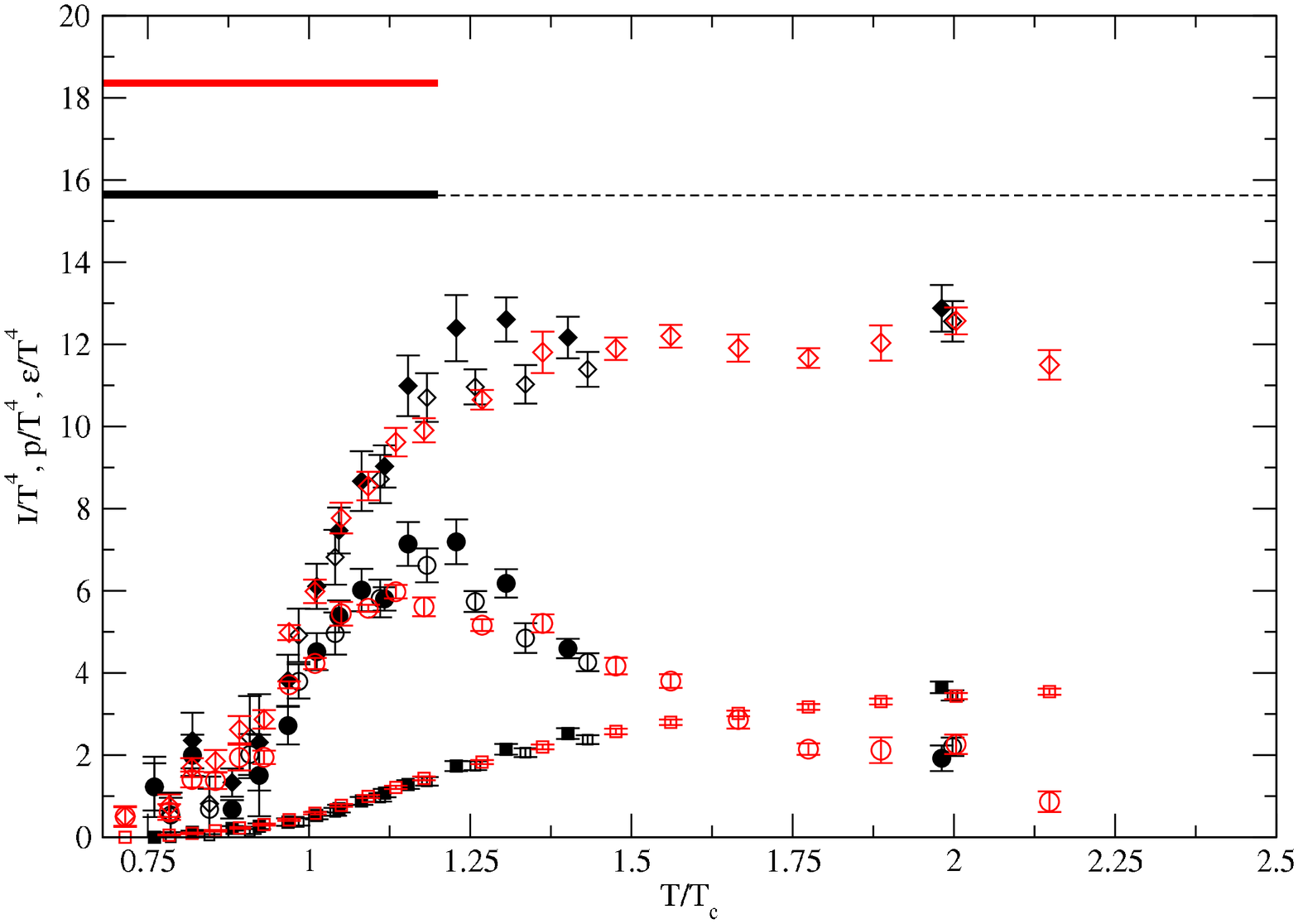}
  \caption{Cut-off dependence of the pressure calculated with the standard 
staggered fermion action on lattices with temporal extent $N_\tau =4$ and
6 in 2-flavor QCD \cite{milc_nt4,milc_nt6} and (2+1)-flavor QCD \cite{aoki} 
(left).  The right hand figure shows energy density, pressure and 
$I=\epsilon - 3p$ for (2+1)-flavor QCD calculated on lattices 
with temporal extent 
$N_\tau =4$ and $6$ with an ${\cal O}(a^2)$ improved staggered fermion action 
\cite{Bernard}.
}
\label{fig:pressure}
\end{figure}

Partly these calculations still have been performed with unimproved fermion
actions \cite{aoki} which makes it difficult to control cut-off effects
and draw quantitative conclusions on the approach to the ideal gas 
limit at high temperature. Nonetheless, these calculations support earlier 
findings on the temperature dependence of the pressure and energy density
in the transition region. In particular, they show that 
the presence of strange quarks has little influence on the thermodynamics
in the vicinity of $T_0$. In Fig.~\ref{fig:pressure} we compare the
recent calculation of the pressure in (2+1)-flavor QCD \cite{aoki}
and earlier results for 2-flavor QCD \cite{milc_nt4,milc_nt6} which both
have been performed with unimproved staggered fermions.
The good agreement in the vicinity of $T_0$ suggests
that the strange quark contribution to the pressure is small. Only for 
$T\; \gsim \; 1.5 T_0$ differences show up; the positive strange quark 
contribution to the pressure in $(2+1)$-flavor
QCD becomes sizeable\footnote{As the relative size of the strange quark 
contribution to thermodynamic quantities changes with temperature it is
obvious that cut-off effects can not be taken into account by rescaling
the pressure or energy density
with a universal, temperature independent factor \cite{aoki}. This underlines 
the need of improved actions for thermodynamic calculations.}.

The good agreement between recent calculations in (2+1)-flavor
QCD, performed with smaller quark masses and smaller lattice spacings
\cite{aoki,Bernard}, and
earlier results \cite{peikert,milc_nt4,milc_nt6} is quite 
reassuring. 
These calculations confirm that thermodynamics in the high 
temperature phase 
is rather insensitive to changes of the quark mass;
a reduction of the light quark masses by almost an order of
magnitude does not lead to drastic changes in the energy density at the
transition point and in the high temperature phase.
Moreover, they 
show that the transition itself is not strongly influenced by
discretization errors, which in the staggered fermion formulation
show up prominently in the distortion of the Goldstone modes; 
reducing $m_q$ and thus the masses of light hadrons as 
well as reducing flavor symmetry breaking effects drastically \cite{aoki} 
does not 
significantly change the energy density at the transition temperature.
The estimate, $\epsilon_c/T_0^4 = 6\pm 2$ \cite{peikert}, is consistent
with the recent calculations in (2+1)-flavor QCD performed with lighter
quark masses. The weak dependence of transition parameters on the quark
mass reflects the importance of numerous heavy resonances that are necessary
to build up the particle and energy density needed for the 
transition to occur \cite{redlich}.

Also the recent studies of the equation of state performed at
vanishing quark chemical potential 
suggest that for physical values of the quark masses the transition to the
high temperature phase of QCD only is a rapid crossover rather than a phase
transition which on finite lattices would be signaled by metastabilities
in bulk thermodynamic observables or the chiral condensate. 
None of the calculations performed so far for QCD with two light
quarks with or without the inclusion of a heavier strange quark gave direct 
evidence for a first order phase transition. 

The missing guidance from any universal behavior in the vicinity of the
transition also influences the determination of the transition temperature
itself. As the transition temperature is determined at quark mass values
which are usually larger than those realized in nature one has to
extrapolate to the physical regime. Depending on the ansatz used for the
quark mass dependence of $T_c$ the extrapolations can
differ by about 5\%. A similar systematic uncertainty arises from the 
calculation of a zero temperature observable that is used to set the
physical units for $T_c$. Thus also the recent determinations of the
transition temperature \cite{Bernard,Schierholz}, which lead to values
$T_c\simeq 170$~MeV still suffer from systematic errors of about 10\%.  


\subsection{Non-zero baryon number density}

\begin{figure}
\epsfig{file=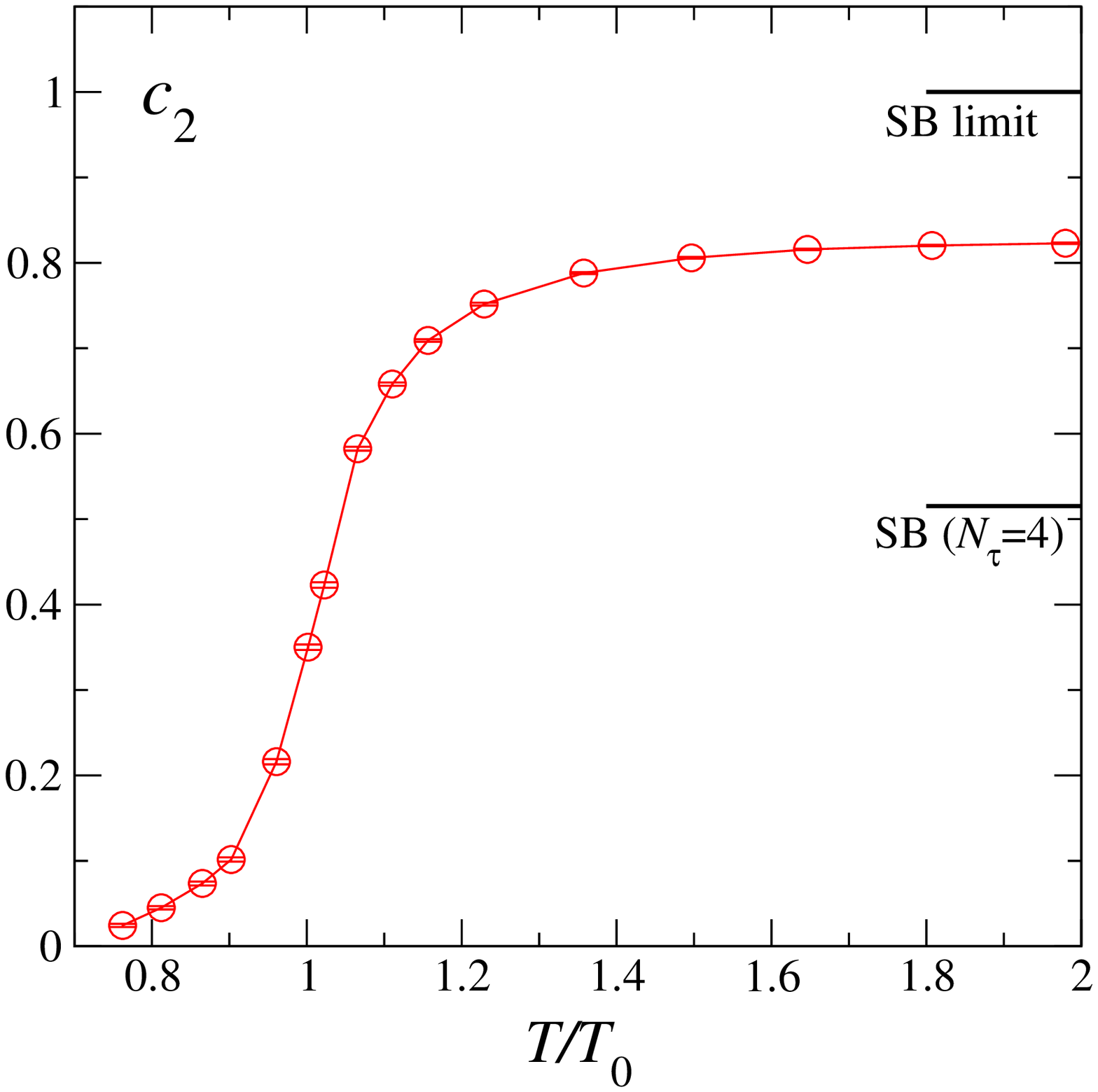,width=4.2cm}
\epsfig{file=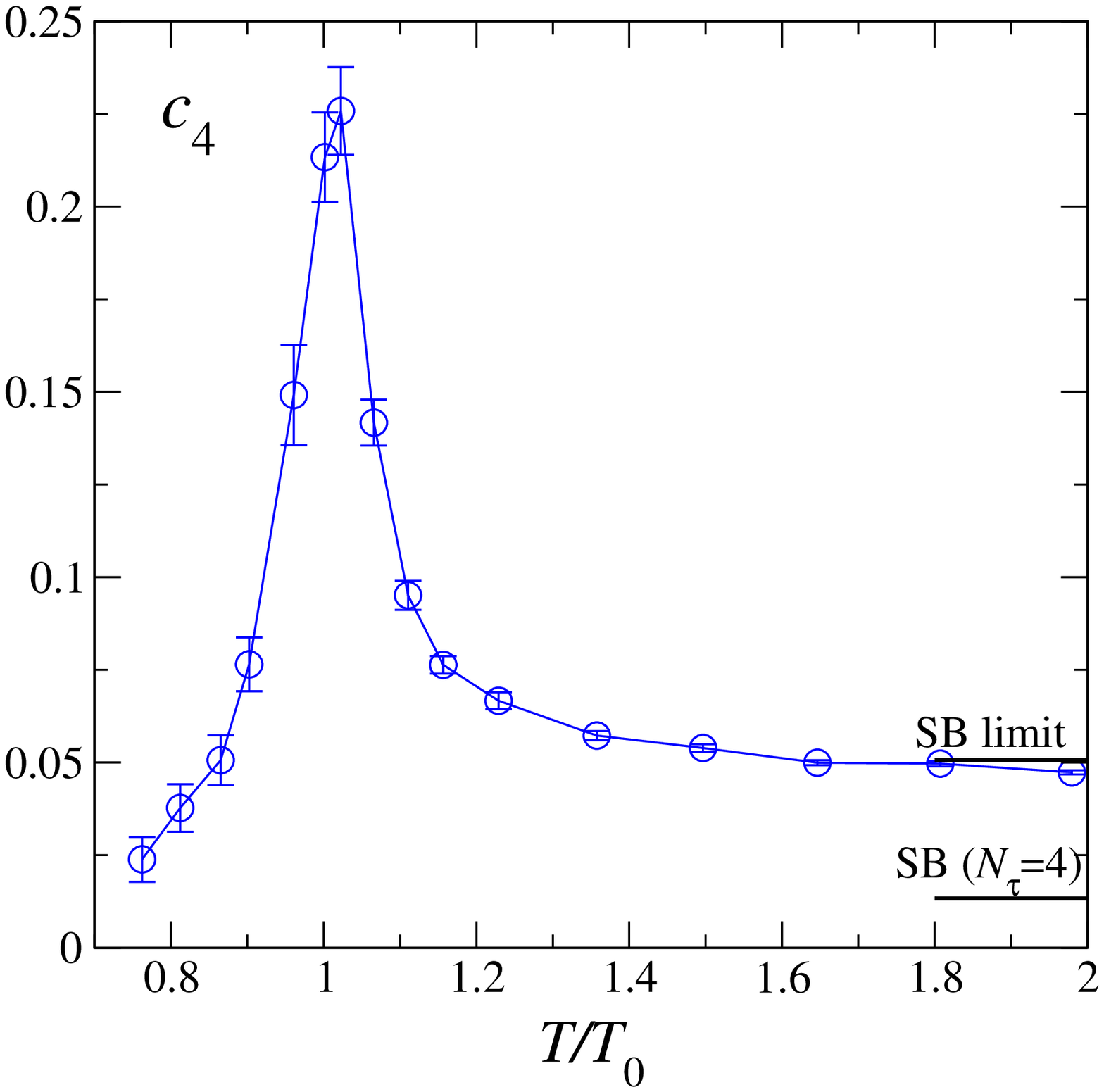,width=4.2cm}
\epsfig{file=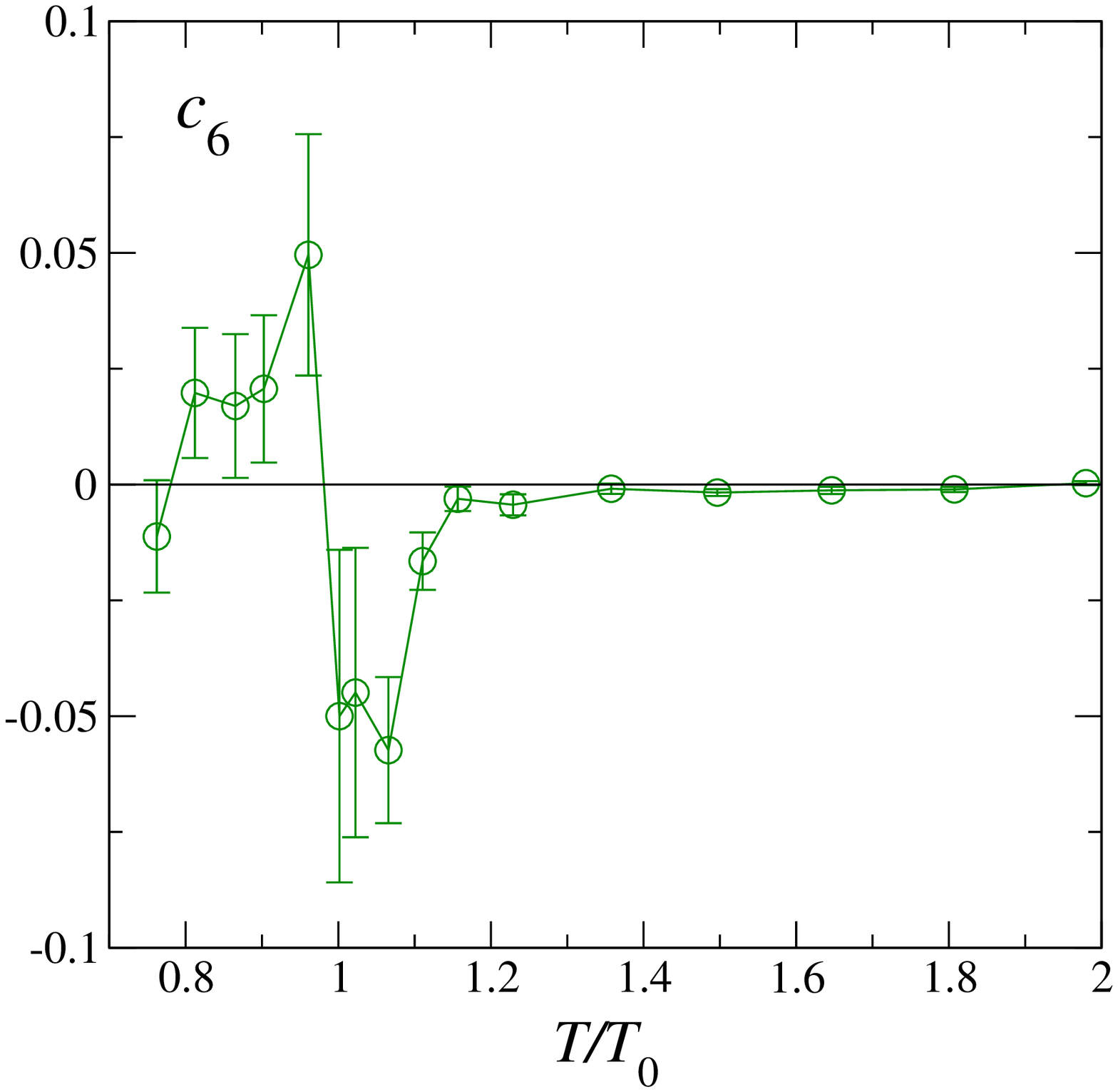,width=4.2cm}
\caption{\label{fig:coefficients} Temperature dependence of expansion
coefficients for $p/T^4$
in 2-flavor QCD and for quark masses corresponding at $T_c$ to a pseudo-scalar 
(pion) mass of about $770$~MeV.
}
\end{figure}

Studies of the QCD equation of state have recently been extended to the
case of non-zero quark chemical potential ($\mu_q$). Calculations of bulk
thermodynamic quantities for $\mu_q > 0$
based on the reweighting approach \cite{Fodoreos},
using the Taylor expansion
of the partition function \cite{Allton2,Allton4,isentropic}, 
as well as analytic continuation of calculations performed with
imaginary values of the chemical potential \cite{lombardo}
show that the 
$\mu_q$-dependent contributions to energy density and pressure are
dominated by the leading order $(\mu_q/T)^2$ correction for 
parameters relevant for the description of dense
matter formed at RHIC ($\mu_q/T\simeq 0.1$). Higher order 
contributions are still only a few percent 
at SPS energies ($\mu_q/T\; \lsim\; 0.6$). 

We will focus in the following
on a discussion of Taylor expansions of the partition function of
2-flavor QCD around $\mu_q=0$.
At fixed temperature and small values of the chemical potential the pressure
may be expanded in a Taylor series around $\mu_q = 0$,
\begin{equation}
{p\over T^4}={1\over{VT^3}}\ln Z = \sum_{n=0}^{\infty} c_n(T,m_q)
\left( \frac{\mu_q}{T} \right)^n \quad ,
\label{Taylorp}
\end{equation}
where the expansion coefficients are given in terms of derivatives of
$\ln Z(T,\mu_q)$, {\it i.e.} $c_n(T,m_q) = \displaystyle{\frac{1}{n!}
\frac{\partial^n \ln Z}{\partial (\mu_q / T)^n}}$.  The series is
even in $(\mu_q/T)$ which reflects the invariance of $Z(T,\mu_q)$ under
exchange of particles and anti-particles. The Taylor series for the
energy density can then be obtained using the relation, 
$(\epsilon -3p)/T^4= T{\rm d}(p/T^4)/{\rm d}T$,
\begin{equation}
\frac{\epsilon}{T^4} = \sum_{n=0}^\infty \left(3 c_n(T,m_q) +
c'_n(T,m_q)\right) \left({\mu_q\over T}\right)^n\quad 
\label{eps}
\end{equation}
with $c'_n(T,m_q) =T {\rm d} c_n(T,m_q)/{\rm d} T$. A similar relation holds
for the entropy density \cite{isentropic}.
The coefficients $c_n(T,m_q)$ calculated for a 
fixed value of the bare quark mass up to $n=6$ 
are shown in Fig.~\ref{fig:coefficients}.

\begin{figure}
\epsfig{file=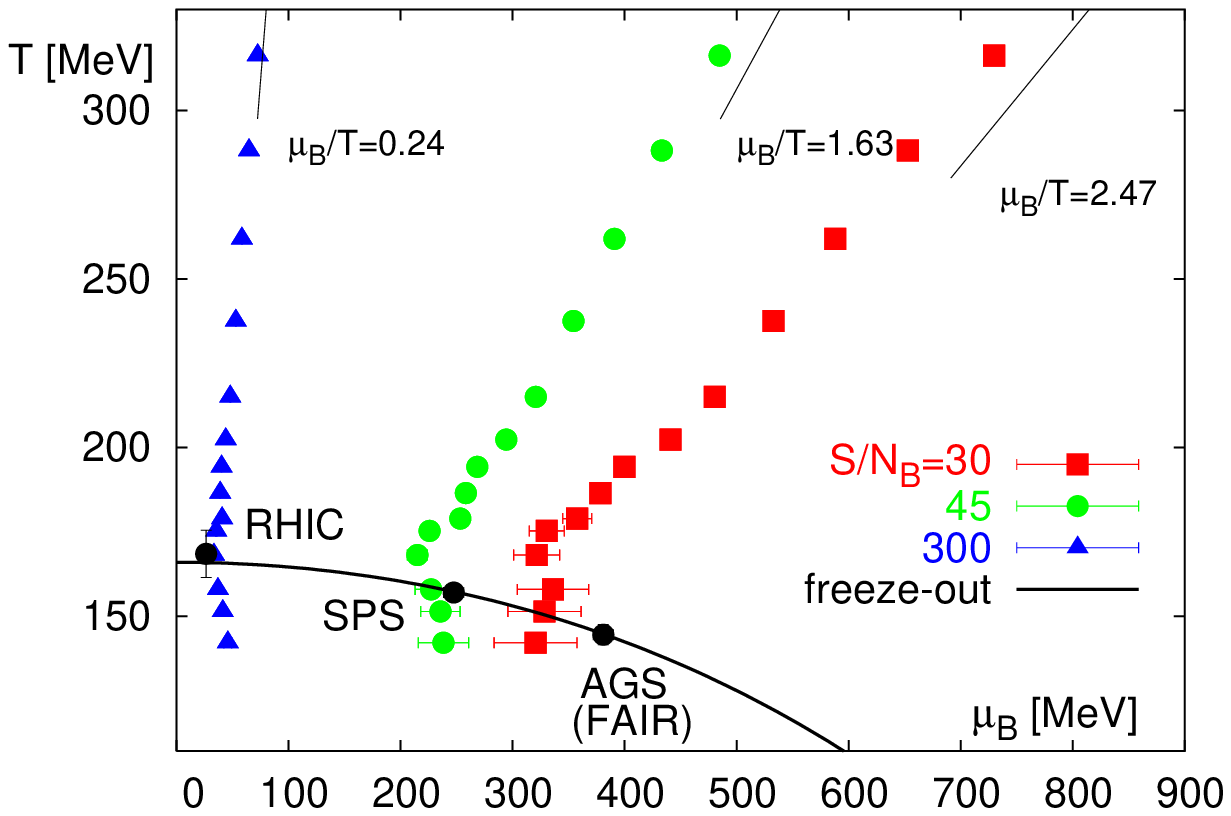,width=7.0cm}\hspace{0.2cm}
\epsfig{file=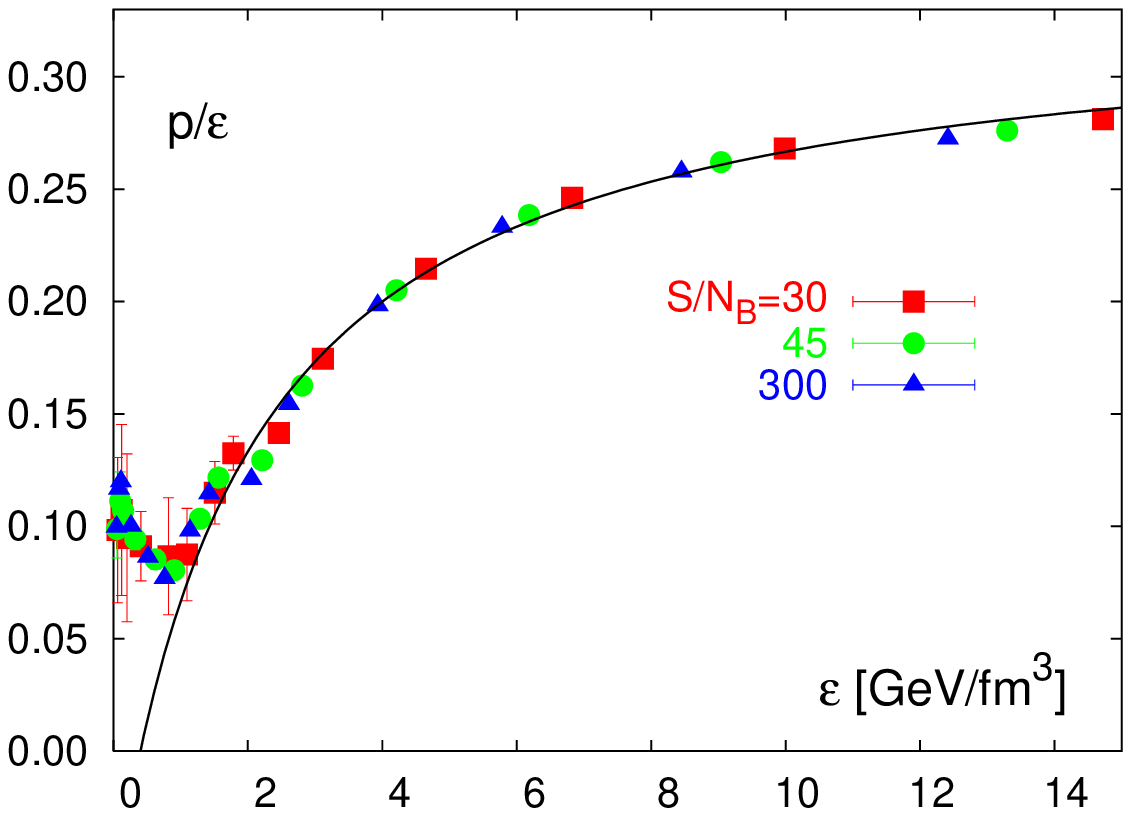, width=7.0cm}
\caption{Equation of state of 2-flavor QCD on lines of constant entropy per 
baryon number. The left hand figure shows three lines of constant $S/N_{\;B}$ 
in the
QCD phase diagram relevant for the freeze-out parameters determined in various
heavy ion experiments. The right hand figure shows the equation of state
on these trajectories using $T_c=175$~MeV to set the
scale. The solid curve in the right hand figure is the
parametrization of the high temperature part of the equation of state
given in Eq.~\protect{\ref{fit}}.}
\label{fig:soft}
\end{figure}

Knowing the dependence of the energy density and the pressure on the
quark chemical potential one can eliminate $\mu_q$ in favor of a 
variable that characterizing the thermodynamic boundary conditions for the 
system under consideration \cite{isentropic}. In the 
case of dense matter created in heavy ion collisions this is a combination
of entropy and baryon number.  Both quantities stay constant during the 
expansion of the system. In Fig.~\ref{fig:soft} we show the
resulting isentropic equation of state as function of temperature as 
well as energy density obtained from a $6^{th}$ order Taylor
expansion of pressure and energy density \cite{isentropic}. 
The three different entropy and baryon number
ratios, $S/N_B = 30,~45$ and $100$, correspond roughly to isentropic
expansions of matter formed at the AGS, SPS and RHIC, respectively. 
It is quite remarkable that $p(\epsilon)$ is to a good approximation
independent of $S/N_B$; for temperatures $T>T_0$, or equivalently
$\epsilon\; \gsim\; 0.8$~GeV/fm$^3$, it is well described by
\begin{equation}
\frac{p}{\epsilon} = \frac{1}{3}\left(1- \frac{1.2}{1+0.5\; \epsilon\; {\rm fm}^3
/{\rm GeV}}\right) \; ,
\label{fit}
\end{equation}
which for large $\epsilon$ agrees with a bag equation of state with
$B^{1/4}\simeq 260$~MeV.

\begin{figure}
\epsfig{file=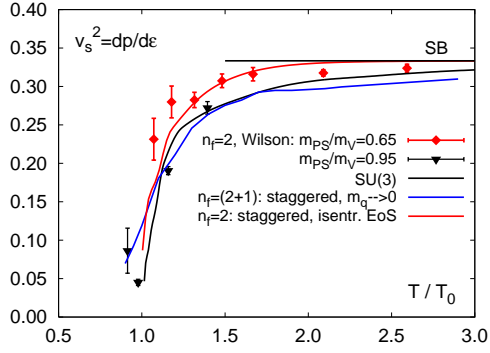,width=7.0cm}
\caption{The velocity of sound in QCD vs. temperature expressed in units
of the transition temperature $T_0$. 
Shown are results from calculations
with Wilson \cite{AliKhan} and staggered fermions \cite{aoki} as well as 
for a pure SU(3) gauge theory \cite{Boyd}.
Also shown is the resulting $v_s^2$ deduced from Eq.~\ref{fit} 
\cite{isentropic}.}
\label{fig:sound}
\end{figure}

The insensitivity of the isentropic equation of state on $S/N_B$ also
implies that the velocity of sound, $v_S^2 = {\rm d} p/{\rm d}\epsilon$
is similar along different isentropic expansion trajectories. In fact,
the parametrization given in Eq.~\ref{fit} suggests that the velocity
of sound approaches rather rapidly the ideal gas value, $v_S^2=1/3$.
In Fig.~\ref{fig:sound} we summarize results for $v_s^2$ obtained in lattice 
calculations for a pure $SU(3)$ gauge theory \cite{Boyd}, 
for 2-flavor QCD with Wilson fermions \cite{AliKhan} and 
($2+1$)-flavor QCD with staggered fermions \cite{aoki}.

\section{The chiral critical point}

Various model calculations \cite{Stephanov} suggest that a second order 
phase transition point (chiral critical point) exists in the QCD phase 
diagram which separates a region of first order phase transitions at high 
baryon number density and low temperatures from a cross-over region at low 
baryon number density and high temperature. Evidence for the existence
of such a critical point may come from lattice calculations at non-zero
quark chemical potential by either determining the location of Lee-Yang
zeroes \cite{Fodor1} or by determining the convergence radius of 
the Taylor series for the logarithm of the partition function which directly 
yields the pressure, $p/T= V^{-1}\ln Z$ \cite{Allton2}.

If there exists a $2^{nd}$ order phase transition point in the QCD phase 
diagram, this could be determined from an analysis of the volume 
dependence of  Lee-Yang zeroes of the QCD partition function.
In any finite
volume zeroes of $Z(V,T,\mu_q)$ only exist in the complex $\mu_q$
plane with ${\rm Im}\mu_q \ne 0$. Only for $V\rightarrow \infty$ 
some of these zeroes may converge to the real axis and will then
give rise to singularities in thermodynamic quantities. The  
relation between phase transitions and zeroes of the partition function
has been exploited using a reweighting technique to extend lattice
calculations at $\mu_q = 0$ to $\mu_q > 0$ \cite{Fodor1}.
Recent results based on this approach \cite{Fodor2} suggest that a critical 
point indeed exists and occurs at $\mu_B=3\mu_q \simeq 360$~MeV.  This 
estimate is about a factor two smaller than earlier estimates \cite{Fodor1}
which have been obtained on smaller lattices and with large quark masses.
This suggests that a detailed analysis of the quark mass and volume 
dependence \cite{ejiri} still is needed to gain confidence in the analysis
of Lee-Yang zeroes.
 
The radius of convergence of the Taylor series is controlled by  
a singular point in the complex $\mu_q$ plane closest to the origin. 
It is related to the location of the critical point
only if this singularity lies on the real axis. A sufficient condition
for this is that all expansion coefficients in the Taylor series are 
positive. For temperatures below the transition temperature at $\mu_q =0$ 
this is indeed the case for all expansion coefficients 
calculated so far. The first coefficient, $c_0$,
just gives the pressure at $\mu_q = 0$ shown in Fig.~\ref{fig:chiral}(right)
and thus is positive for all temperatures.
This also is the case for  $c_2$, which
is proportional to the quark number susceptibility at $\mu_q = 0$
\cite{Gottlieb},
\begin{equation}
\frac{\chi_q}{T^2} = \frac{\partial^2 p/T^4}{\partial (\mu_q/T)^2} =
\sum_{n=0}^{\infty}d_n \left( \frac{\mu_q}{T} \right)^n \quad {\rm with}
\quad d_n=(n+2)(n+1)c_{n+2} \; . 
\label{chiq}
\end{equation}
As can be seen in Fig.~\ref{fig:pressure} also the next-to-leading
order coefficient, $c_4$, is strictly positive.

A new feature shows up in the expansion coefficients at ${\cal O}(\mu^6)$.
The coefficient $c_6$ is positive only below $T_0$ and changes sign in its
vicinity. If this pattern persists for higher order
expansion coefficients one may conclude that the 
irregular signs of the expansion coefficients for $T>T_0$ suggest that
the radius of convergence of the Taylor series is not related to critical
behavior at these temperatures, whereas it determines a critical point
for $T<T_0$ if also the higher order expansion coefficients stay positive.

Ratios of subsequent expansion coefficients provide an estimate
for the radius of convergence of the Taylor expansion,
\begin{equation}
\rho (T)=\lim_{n\to\infty}\rho_n\equiv
\lim_{n\to\infty}\sqrt{\left\vert {c_n\over c_{n+2}}\right\vert} =
\lim_{n\to\infty}\sqrt{\left\vert {d_n\over d_{n+2}}\right\vert} \quad .
\label{convergence}
\end{equation}
The expansion coefficients $d_n$ for the quark number susceptibility have
been analyzed recently for unimproved staggered fermions \cite{gavai} up
to $n=6$. It has been shown that an accurate determination of these expansion 
coefficients requires large physical volumes. Based on a finite volume
analysis the radius of convergence 
has been estimated from the Taylor series of the quark number
susceptibility to be $\mu_B \simeq 180$~MeV \cite{gavai}. As the expansion
coefficients $d_n$ are directly related to the expansion coefficients $c_n$
of the pressure the radius of convergence coincides in the limit $n\rightarrow
\infty$. For finite $n$, {\it i.e.} $n\simeq 6$, estimates based on the 
Taylor series for the pressure will be about 30\% higher than those based
on the Taylor series for susceptibilities. 
 
\begin{figure}
\epsfig{file=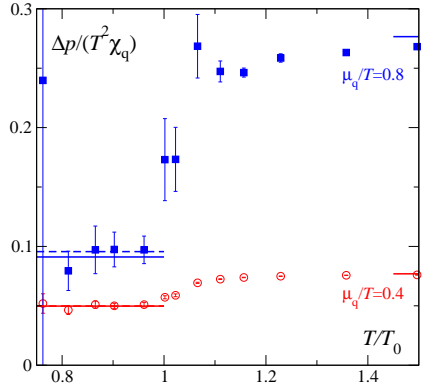,width=5.5cm}
\caption{\label{fig:ratios_coefficients} 
The dimensionless ratio
$\Delta p/T^2 \chi_q$ for two values of the chemical potential $\mu_q /T$.
Horizontal lines show the expected results for a hadronic resonance gas and an
ideal quark-antiquark gas below and above $T_0$, respectively.
}
\end{figure}

While the quark number susceptibility is expected to diverge at the 
chiral critical point, the pressure, of course, will stay finite. 
Although $\chi_q$ rises rapidly with increasing $\mu_q/T$, this is 
partly due to the rapid increase of the pressure itself. 
A quantity reflecting the relative
magnitude of fluctuations is given by the ratio,
\begin{equation}
\frac{\Delta p}{T^2 \chi_q} =
\frac{1}{2}\;  \biggl( {\mu_q \over T}\biggr)^2 {
1 + \frac{c_4}{c_2} \left( {\mu_q \over T}\right)^2
+ \frac{c_6}{c_2} \left( {\mu_q \over T}\right)^4  + 
{\cal O}\left( \left( {\mu_q \over T}\right)^6\right)
\over
1 + 6 \frac{c_4}{c_2} \left( {\mu_q \over T}\right)^2
+ 15 \frac{c_6}{c_2} \left( {\mu_q \over T}\right)^4 +
{\cal O}\left( \left( {\mu_q \over T}\right)^6\right)
}
\label{ratio}
\end{equation}
This quantity only depends on ratios of Taylor expansion coefficients,
and should vanish at a $2^{nd}$ order phase transition point. It is
shown in Fig.~\ref{fig:ratios_coefficients}. As
can be seen the ratio rises rapidly across $T_0$ and approaches the ideal gas
value. It, however, does not show any sign of a drop in the vicinity of $T_0$
that could be taken as evidence for the existence of a second order
transition.  For this reason it has been argued in Ref.~\cite{Allton4}
that conclusions on the radius of convergence drawn from estimators $\rho_n$
may be premature at least for the quark masses used in this calculations.
As these masses were significantly larger than those used in Ref.~\cite{gavai}
both results are not necessarily in contradiction. 
Like in the case of studies of Lee-Yang zeroes a more detailed
analysis of the quark mass dependence also is needed in this case.

\begin{theacknowledgments}
This manuscript has been authored under contract number
DE-AC02-98CH1-886 with the U.S. Department of Energy.
Accordingly,
the U.S. Government retains a non-exclusive, royalty-free license to
publish or reproduce the published form of this contribution, or allow
others to do so, for U.S.~Government purposes.
\end{theacknowledgments}

\end{document}